\documentclass[sigconf]{acmart}
\usepackage{booktabs} 
\usepackage{makecell,multirow,diagbox,underscore,graphicx,enumerate,supertabular,lscape}
\usepackage{rotating} 
\usepackage{longtable}

\copyrightyear{2017} 
\acmYear{2017} 
\setcopyright{acmcopyright}
\acmConference[UrbanGIS'17]{UrbanGIS'17:3rd ACM SIGSPATIAL Workshop on Smart Cities and Urban Analytics }{Nov. 2017}{Redondo Beach, CA, USA}
\acmPrice{15.00}\acmDOI{10.1145/3152178.3152182}
\acmISBN{978-1-4503-5495-0/17/11}

\begin{document}
\title{Quantitative Comparison of Open-Source Data for Fine-Grain Mapping of Land Use}
  
\author{Xueqing Deng}
\affiliation{%
  \institution{Electrical Engineering and Computer Science\\ University of California, Merced}
}
\email{xdeng7@ucmerced.edu}

\author{Shawn Newsam}
\affiliation{%
  \institution{Electrical Engineering and Computer Science \\
  University of California, Merced}
}
\email{snewsam@ucmerced.edu}


\begin{abstract}

This paper performs a quantitative comparison of open-source data available on the Internet for the fine-grain mapping of land use. Three points of interest (POI) data sources--Google Places, Bing Maps, and the Yellow Pages--and one volunteered geographic information data source--Open Street Map (OSM)--are compared with each other at the parcel level for San Francisco with respect to a proposed fine-grain land-use taxonomy. The sources are also compared to coarse-grain authoritative data which we consider to be the ground truth. Results show limited agreement among the data sources as well as limited accuracy with respect to the authoritative data even at coarse class granularity. We conclude that POI and OSM data do not appear to be sufficient alone for fine-grain land-use mapping.

\end{abstract}

%
%



\maketitle
%

\section{Introduction}

Land use information plays an important role in urban planning and can inform city design and utility distribution \cite{Jamal13}. Land use refers to the function of the land, which is shaped by human activities \cite{Mao:2016:EMP:3007540.3007549}, such as education, retail, etc. It is different from land cover, such as vegetation, built-up areas, etc., which is determined by the land's physical attributes. Remote sensing can be used to determine land cover; mapping land use, however, is much more challenging. The most accurate method for assessing land use has traditionally been through surveys. This is labor intensive and time consuming, and is soon outdated. More automated methods for mapping land use are needed.

Land use and land cover are often treated together. There are many combined land use and land cover (LULC) classification systems but they typically blur the distinction between the two super classes and tend to be relatively coarse grain. The European Urban Atlas (UA) project\footnote{https://www.eea.europa.eu/data-and-maps/data/urban-atlas\#tab-gis-data} is one example. UA provides consistent LULC data for urban zones with more than one hundred thousand people across Europe. It has a well defined mapping methodology and a hierarchical taxonomy of 17 urban and 10 rural classes. To our knowledge, no LULC mapping effort at this scale and even this relatively coarse granularity exists in the United States. The evaluation performed in this paper is a step towards an automated method for fine-grain LU classification in the United States and beyond. Significantly, we undertake the key step in this paper of establishing a LU class taxonomy that is finer grained than any previous system and whose classes are distinct from LC. 

A range of techniques have been developed for automated LULC classification, including using remote sensing imagery \cite{Cheng15}, social media \cite{Zhu:2015:LUC:2820783.2820851}, cell phone data \cite{Toole12}, and points of interest \cite{Yao17}, or combinations of sources \cite{Liu17}. Classification based on remote sensing imagery has perhaps the longest history but the resulting products tend to confuse land use and land cover and are coarse grain \cite{Elhadi14,rs1030330,SAADAT2011608}. More recently, ground-level imagery has been investigated for LU classification \cite{Zhu:2015:LUC:2820783.2820851,Leung:2012:EGI:2390790.2390794}. The different and close-up perspective of this imagery has the potential to detect function, particularly indoors. However, this approach is limited by the availability of georeferenced ground-level images.

Points of interest (POI) data is a particularly promising source of data for LU mapping. It is readily available online, often through well-developed application programming interfaces (APIs), and typically consists of geographic coordinates and a specific type or category such as restaurant, bank, etc. Previous work has investigated POI data for LU mapping \cite{Yao17,jiang2015mining} or well as other applications such as mapping population \cite{Mohamed14}. A key challenge in evaluating LU classification is the lack of ground truth. POI data has therefore also been used as reference set \cite{Mao:2016:EMP:3007540.3007549} although its validity as ground truth is not clear.

Another source of data for land use mapping is volunteered geographic information (VGI), a term introduced by Goodchild \cite{goodchild2007citizens} in 2007 to refer to geographic data that is created, assembled, and disseminated voluntarily by individuals. Open Street Map (OSM) is perhaps the most well-known example of VGI. The LU information available in OSM has been compared with authoritative LU data \cite{arsanjani2015quality} but this study was limited to Germany where OSM data is more complete. OSM is much less complete outside Europe, particularly in the United States \cite{TGIS:TGIS12037} and in China, where 94\% of the country had little or no data as of 2014 \cite{Zheng2014}.

A wide range of open-source data has been used for mapping LU. However, it is not clear how these sources differ. We therefore undertake the first comparison, to our knowledge, of these different sources. We do this with respect to a new, fine-grain LU class taxonomy which we introduce. We focus on POI and VGI data as these seem to be the most promising sources for LU mapping. We compare the sources to each other as well as to a coarse-grain authoritative LU map.

We summarize our contributions as follows: 
\begin{itemize}

\item We introduce a new, fine-grain LU class taxonomy based on the American Planning Association's Land Based Classification Standard \cite{LBCS}. This taxonomy characterizes function. It is hierarchical with 9 level-one classes, 47 level-two classes, and 159 level-three classes. We refer to this as the LBCS LU classes. The LBCS hierarchy relevant to this study is shown in the first four columns of table \ref{tab:LBCS1} and the first two columns of table \ref{tab:LBCS2}.

\item We compare three POI sources, Google Places, Bing Maps, and the Yellow Pages, and one VGI source, OSM, with respect to mapping the LBCS classes at the parcel footprint level for the city of San Francisco. We compare the sources to each other as well as to a coarse-grain authoritative LU map. This is the first time, to our knowledge, that such a number of sources has been compared.

\end{itemize}

\section{Overview of the Study}

A data source can be deficient in a number of ways for mapping LU with respect to a particular class taxonomy over a given geographic region. The source's classes might not align with the target classes. That is, classes could be missing or not at same taxonomic level. The location information of the data might not be accurate. And, the spatial coverage could be sparse. Ground truth would allow a data source to be quantitatively assessed along these three dimensions. No ground truth exists for our LBCS classes and so we instead compare our sources to each other to provide insight into their individual deficiencies.

We first align each source with the LBCS taxonomy. This is a difficult undertaking since the sources were not created for LU classification. They also differ significantly among each other with respect to their taxonomic structure. We then use the geographic locations of the source data to assign LBCS classes to the parcel footprints. This allows us to assess the spatial and taxonomic coverage of the individual data sources. We also quantitatively compare them at the footprint scale.

We do have access to coarse-grain authoritative LU information at the parcel footprint level for the study region. We compare each of the sources with this information.

\section{Datasets}
This section describes the datasets used in the study. We download POI data from Google Places \cite{Google} using its API, from Bing Maps \cite{Bing} using its API, and from the Yellow Pages website\footnote{https://www.yellowpages.com/}. We download OSM points and polygons in ESRI shape format from QGIS \footnote{http://www.qgis.org/en/site/}. Finally, we download the authoritative LU data including the parcel footprints from DataSF \cite{SFopendata}. Thus our dataset can be divided into three categories, POI, OSM features including points and polygons, and authoritative data.

\vspace{-5pt}
\subsection{POI}
We obtain 55,126 records from the Google Places API for 74 relevant place types (out of 91) for San Francisco City. Examples of relevant place types include ``bank'', ``museum'', and ``restaurant''. We obtain 7,601 records from the Bing Maps API using 39 relevant entity types (out of 69). Examples of entity types include ``shopping'', ``hotel'', and ``ATM''. We obtain 42,183 records from the Yellow Pages website by searching for the 74 Google Places place types. We wrote our own script to parse the Yellow Pages search results.

\vspace{-5pt}
\subsection{OSM}
OSM data can be accessed either through its own API or third-party open-source tools. We used QGIS, an application which can download OSM data in XML format and convert it to ESRI shapefiles. We extract 31,784 points and 161,285 polygons in a bounding box of San Francisco City. Unlike the POI data, OSM attributes do not have a fixed set of values. Instead, contributors are free to use any description and even create new attributes. Examples of OSM attributes for San Francisco include ``land use'', ``building'', ``railway'', and ``shop''. The values assigned to these attributes often overlap. Of the 193,069 records downloaded from OSM, only 10,439 have non-empty relevant attribute values.

\vspace{-5pt}
\subsection{Authoritative Data}
We download the parcel footprints as ESRI shapefiles for San Francisco from DataSF \cite{SFopendata}. There are a total of 245,003 parcel polygons. This data also includes coarse-grain LULC labels for each footprint. This flat taxonomy contains 12 classes, 10 of which are relevant to land use. These 10 classes are listed in table \ref{tab:DataSFClasses}.

\begin{table}[t]
\caption{Land use classes from DataSF \vspace{-10pt}}
\begin{tabular}{p{2cm}|p{5cm}} 
\hline  
\hline  
Name & Description \\
\hline
CIE & Cultural, Institutional, Educational\\
\hline
MED & Medical\\
\hline
MIPS & Office (Management, Information, Professional Services)\\
\hline
MIXED & Mixed Use (Without Residential)\\
\hline
MIXRES & Mixed Use (With Residential)\\
\hline
PDR & Industrial (Production, Distribution, Repair)\\
\hline
RETAIL/ENT & Retail, Entertainment\\
\hline
RESIDENT & Residential\\
\hline
VISITOR & Hotels, Visitor Services\\
\hline
VACANT & Vacant\\
\hline
\hline
\end{tabular} 
\label{tab:DataSFClasses}
\end{table}  

\vspace{-5pt}
\section{Methodology}
\begin{figure}
\setlength{\abovecaptionskip}{0.1cm}
\setlength{\belowcaptionskip}{-0.cm}
\includegraphics[width=\linewidth]{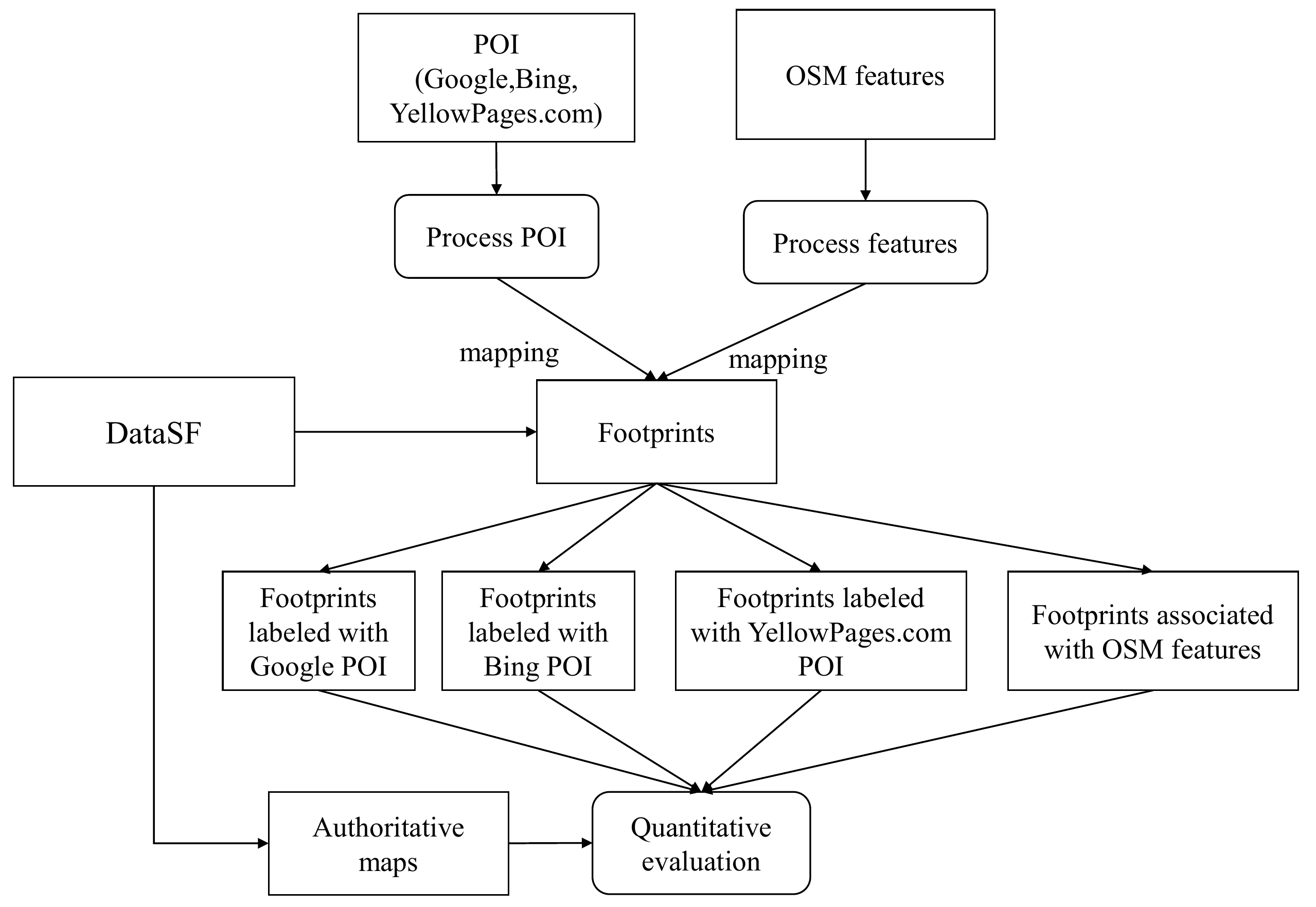}
\caption{Flowchart of the study\vspace{-10pt}}
\label{fig:study-flowchart}
\end{figure}

This section describes how the POI and OSM data is used to assign LBCS classes to the parcel footprints. Figure \ref{fig:study-flowchart} shows a flowchart of the overall study.

\vspace{-2pt}
\subsection{Mapping POI Data}
We first manually align the POI data with the LBCS classes. This alignment is shown in tables \ref{tab:LBCS1} and \ref{tab:LBCS2}. The first few columns of these tables show the target LBCS classes. The last three columns show the assignment of the Google Places, Bing Maps, and Yellow Pages POI data.

\begin{figure}
\setlength{\abovecaptionskip}{0.1cm}
\setlength{\belowcaptionskip}{-0.cm}
\includegraphics[width=\linewidth]{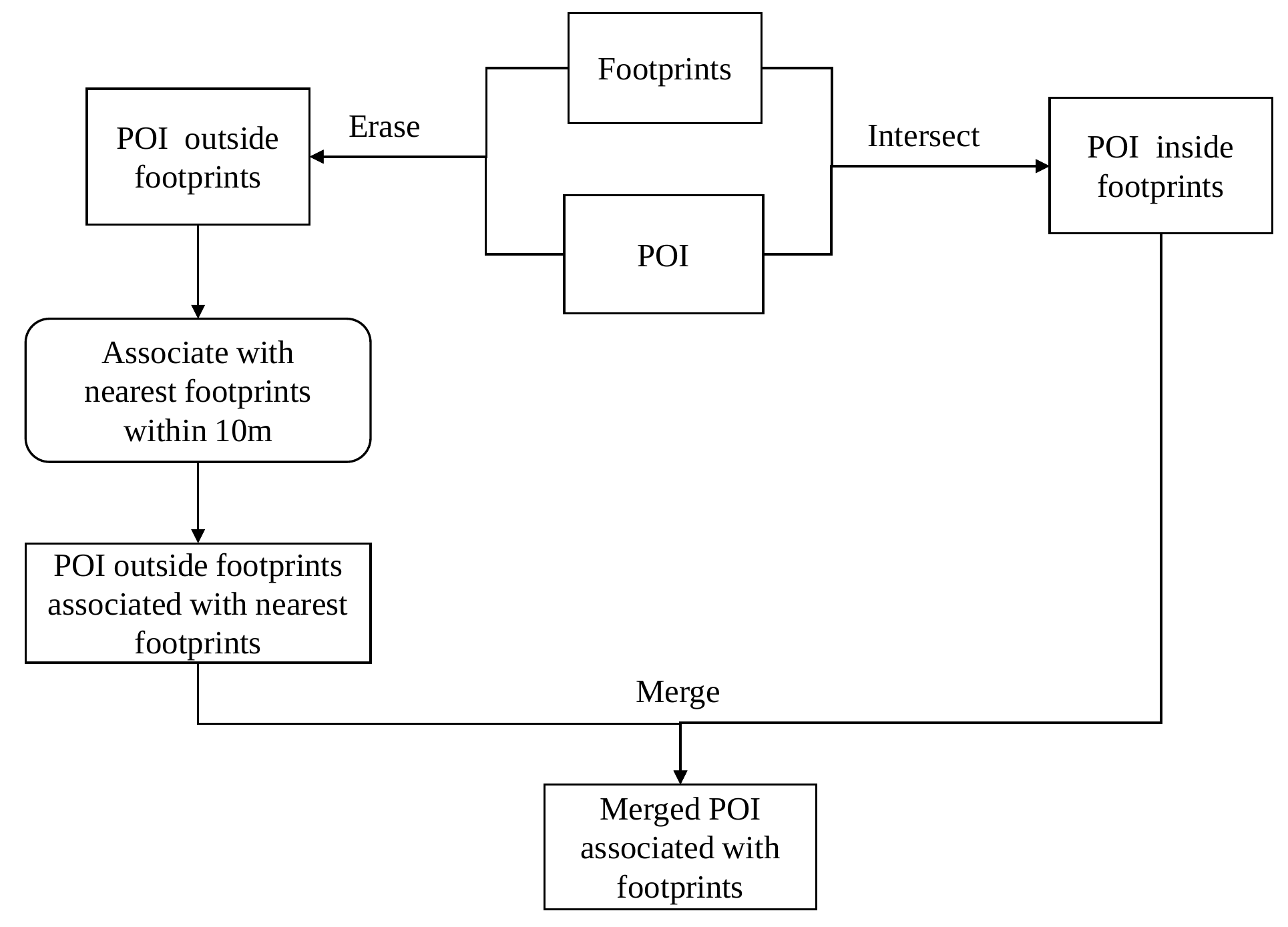}
\caption{Labeling parcel footprints with POI data \vspace{-10pt}}
\label{fig:mapping-POIs}
\end{figure}

Once we have associated an LBCS class with each POI, we label the footprints using the workflow shown in figure \ref{fig:mapping-POIs}. The POI's LBCS class is assigned to whichever footprint the POI falls in. If the POI does not fall in any footprint, we assign its class to the nearest footprint in a 10m radius. POIs that do not fall within 10m of a footprint are ignored. Note that a footprint can thus be labeled with multiple LBCS classes by a single source. This makes sense because a parcel can have more than one land use.

\vspace{-2pt}
\subsection{Mapping OSM Data}
We again first manually align the OSM data with the LBCS classes. The challenge here is that OSM data does not have a fixed set of attributes (keys) and values. We therefore first identify a set of commonly used keys relevant to our application. This includes keys such as ``amenity'', ``building'', and ``land use''. We then identify a set of relevant values for these keys and associate them with the LBCS classes. Table \ref{tab:OSM-alignment} shows the alignment between OSM keys and key values and LBCS classes.

OSM data consists of points and polygons. Points are used to label footprints the same way as the POI data above. Polygons are used to label footprints using shape intersection. Labels are assigned if there is a non-zero intersection between the OSM polygon and a footprint.
\newline
\newline

\begin{figure}
\setlength{\abovecaptionskip}{0.2cm}
\setlength{\belowcaptionskip}{-0.cm}
\includegraphics[width=\linewidth]{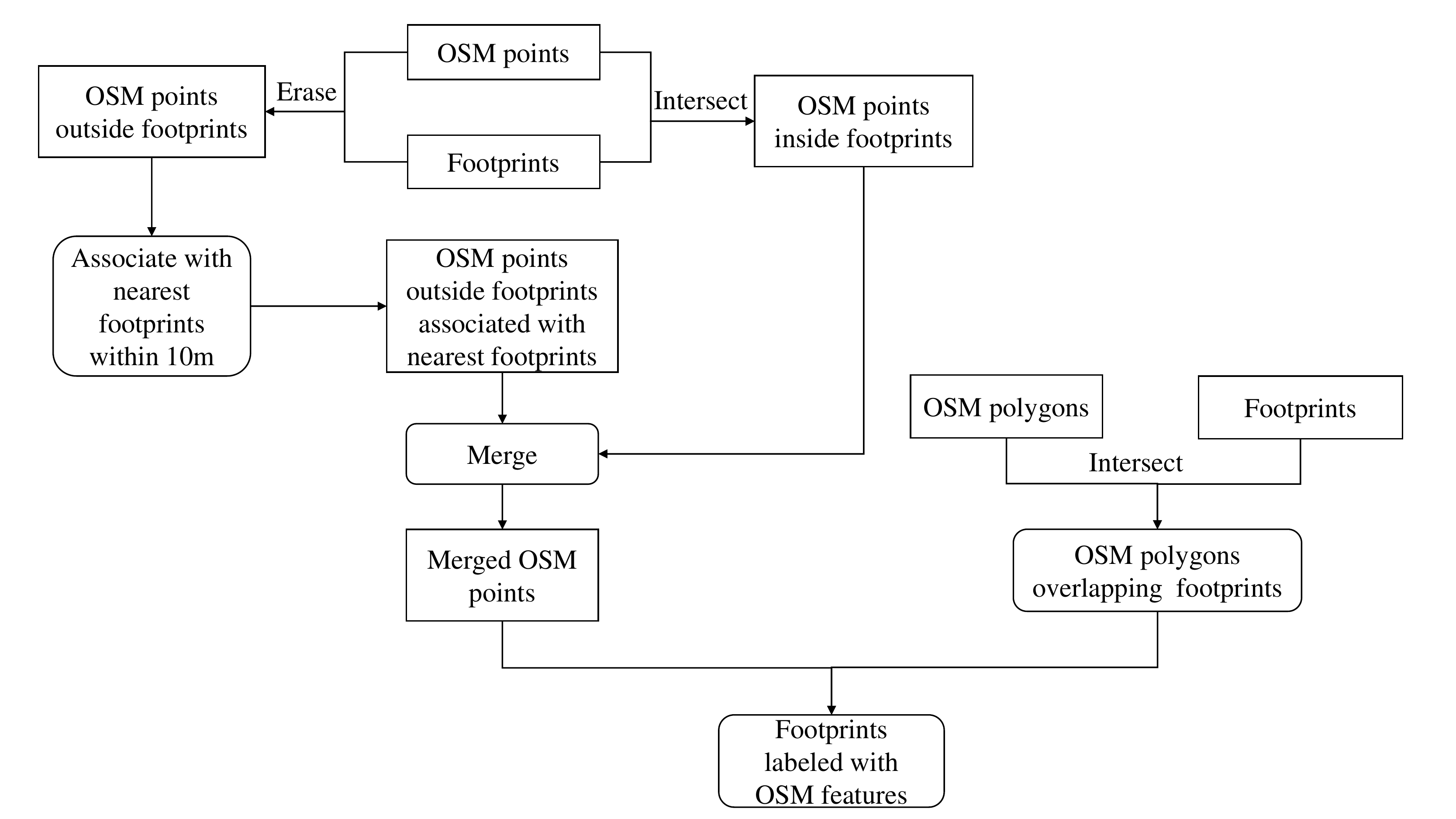}
\caption{Labeling parcel footprints with OSM data \vspace{-10pt}}
\label{fig:mapping-OSM}
\end{figure}

\begin{center}
\setlength{\abovecaptionskip}{0.2cm}
\setlength{\belowcaptionskip}{-0.05cm}
\tablefirsthead{%
\hline
\multicolumn{1}{|c|}{OSM Key} &
\multicolumn{1}{c}{OSM Key Value} &
LBCS \\
\hline}
\tablehead{%
\hline
\multicolumn{3}{|l|}{\small\sl continued from previous page}\\
\hline
\multicolumn{1}{|c}{OSM Key} &
\multicolumn{1}{c}{OSM Key Value} &
LBCS \\
\hline}
\tabletail{%
\hline
\multicolumn{3}{|r|}{\small\sl continued on next page}\\
\hline}
\tablelasttail{\hline}
\tablefirsthead{%
\hline
\multicolumn{1}{|c}{OSM Key} &
\multicolumn{1}{c}{OSM Key Value} &
LBCS \\
\hline}
\tablehead{%
\hline
\multicolumn{3}{|l|}{\small\sl continued from previous page}\\
\hline
\multicolumn{1}{|c}{OSM Key} &
\multicolumn{1}{c}{OSM Key Value} &
LBCS \\
\hline}
\tabletail{%
\hline
\multicolumn{3}{|r|}{\small\sl continued on next page}\\
\hline}
\tablelasttail{\hline}
\tablefirsthead{%
\hline
\multicolumn{1}{|c|}{OSM Key} &
\multicolumn{1}{c|}{OSM Key Value} &
LBCS \\
\hline}
\tablehead{%
\hline
\multicolumn{3}{|l|}{\small\sl continued from previous page}\\
\hline
\multicolumn{1}{|c|}{OSM Key} &
\multicolumn{1}{c|}{OSM Key Value} &
LBCS \\
\hline}
\tabletail{%
\hline
\multicolumn{3}{|r|}{\small\sl continued on next page}\\
\hline}
\tablelasttail{\hline}
\bottomcaption{Alignment of OSM keys/values to LBCS classes}
\label{tab:OSM-alignment}
\begin{supertabular}{|p{1cm}|p{15.28em}|p{1cm}|}

\hline
 \multirow{13}{1cm}{amenity  (point)}
 &bicycle repair station, car wash, clothes stores, corner market, fuel, grocery, market place & 2100 \\

 \cline{2-3}         & bank\&atm & 2200 \\
\cline{2-3}          & car rental & 2300 \\
\cline{2-3}          & animal shelter, embassy, laundry, pet grooming, post office, veterinary, conference center & 2400 \\
\cline{2-3}          & bar, caf\'{e}, fast food, night club, restaurant & 2500 \\
\cline{2-3}          & bicycle parking, bus station, parking & 4100 \\
\cline{2-3}          & library & 4200 \\
\cline{2-3}          & arts center, cinema, music venue & 5100 \\
\cline{2-3}          & gym   & 5300 \\
\cline{2-3}          & college, kindergarten, music school, university & 6100 \\
\cline{2-3}          & fire station, police & 6400 \\
\cline{2-3}          & clinic, community center, hospital, dentist, doctors, doctors office, nursing home & 6500 \\
\cline{2-3}          & place of worship & 6600 \\
\hline
    \multirow{3}{1cm}{building (point)} &   apartment, house, residential & 1100 \\      
\cline{2-3}          & commercial, retail & 2000 \\
\cline{2-3}          & school & 6100 \\
\hline
\multirow{3}{1cm}{landuse (point)} & residential & 1100 \\
\cline{2-3}
& commercial, retail & 2000\\
\cline{2-3}
& recreation & 5000\\
\hline
 \multirow{13}{1cm}{amenity  (polygon)}  & commercial & 2000\\
 \cline{2-3}
 &  car wash, fuel,  market place, pharmacy & 2100 \\

 \cline{2-3}         & bank\&atm & 2200 \\
\cline{2-3}          & car rental, boat rental & 2300 \\
\cline{2-3}          & animal shelter, embassy,  post office, veterinary, conference center & 2400 \\
\cline{2-3}          & bar, caf\'{e}, fast food, night club, restaurant & 2500 \\
\cline{2-3}          & bicycle parking, bus station, parking & 4100 \\
\cline{2-3}          & library, studio & 4200 \\
\cline{2-3}          & arts center, cinema,theater & 5100 \\
\cline{2-3}          & college, kindergarten, school, preschool, university & 6100 \\
\cline{2-3}          & fire station, police & 6400 \\
\cline{2-3}          & clinic, community center, hospital, dentist, doctors, doctors office, nursing home & 6500 \\
\cline{2-3}          & place of worship & 6600 \\
\hline
    \multirow{6}{1cm}{building (polygon)} & apartment, house, residential & 1100 \\
\cline{2-3}       &hotel& 1300 \\    
\cline{2-3}          & commercial, retail & 2000 \\
\cline{2-3}          &train_station &4100\\
\cline{2-3}          &library, museum &4200\\
\cline{2-3}          & school, college, kindergarten, university & 6100 \\
\cline{2-3}          & hospital & 6500\\
\cline{2-3}          & church & 6600\\
\hline
\multirow{3}{1cm}{landuse (polygon)} & residential & 1100 \\
\cline{2-3}
& commercial, retail & 2000\\
\cline{2-3}
& recreation & 5000\\
\hline
\end{supertabular}
\end{center}

\vspace{-4pt}
\section{Quantitative evaluation}
This section presents our quantitative evaluation. This includes comparing the different sources with each other as well as with the authoritative data.
\vspace{-6pt}
\subsection{Spatially Valid Data}

\begin{figure}
\setlength{\abovecaptionskip}{0.3cm}
\setlength{\belowcaptionskip}{-0.2cm}
\includegraphics[width=\linewidth]{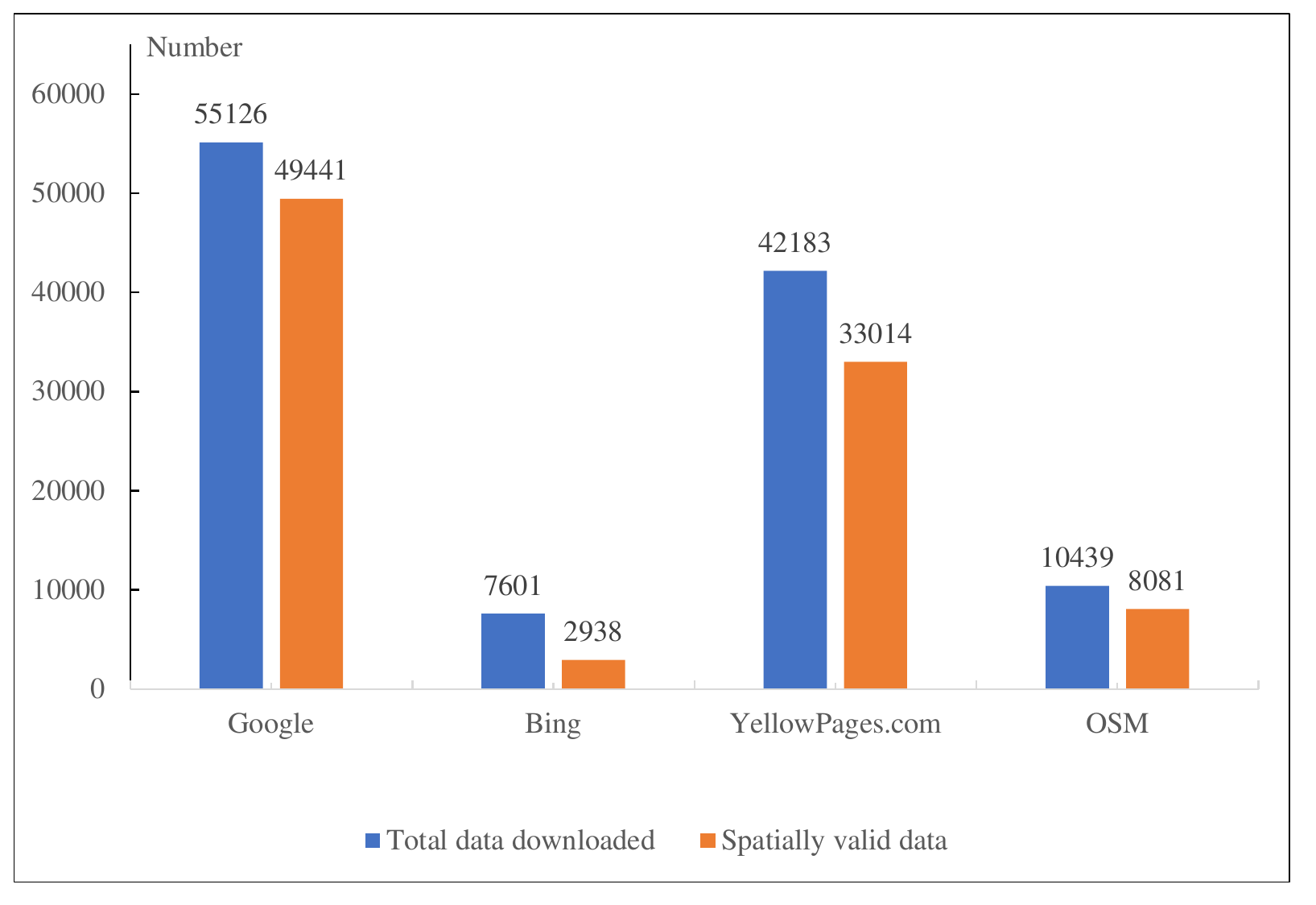}\vspace{-10pt}
\caption{Records before and after mapping to footprints \vspace{-15pt}}
\label{fig:valid-data}
 
\end{figure}

Figure \ref{fig:valid-data} shows the number of records that remain after spatially mapping the data to the footprints. The reduction in valid (that within 10m of a parcel) data is likely due to several factors. First, is simple errors in the location information. Also, some of the records correspond to features which do not fall within footprints such as taxi stands and bus stops. Significantly, none of the data sources contains more than 50,000 valid records. This means that less than 20\% of our target set of 245,003 footprints can be labeled with any one source. Our first finding is thus that both the POI and OSM datasets are sparse at the footprint scale.

Columns 2 through 5 of table \ref{tab:quan-results} show the breakdown of valid data for each of the sources with respect to the LBCS hierarchy.

 \vspace{-2pt}
\subsection{Pairwise Comparisons of Data Sources}

We perform pairwise comparisons between the data sources to determine their level of agreement. High levels of agreement between multiple data sources can be an indication of good accuracy of each especially in the absence of ground truth.

Columns 2 through 5 of table \ref{tab:quan-results} show the number of parcels labeled by each of the datasets for each class. Columns 6 through 11 show the agreement between pairs of data sources and column 12 shows the agreement between all of the sources. These columns indicate the number of parcels labeled consistently by the combinations of sources (their agreement). The number in parentheses is this number divided by the total number parcels labeled by either or both datasets (intersection over union) reported as a percentage. For example, according to the first row, Google Places labeled 586 parcels and Bing Maps labeled 167 parcels with class 1000. 113 of these are in agreement. This represents 17.66\% of the parcels labeled as class 1000 by either Google or Bing or by both.

We make the following observations based on the results in table \ref{tab:quan-results}. Google Places and the Yellow Pages tend to have the highest agreement. This is as high as 49.33\% agreement at level-one of the class hierarchy for class 2000 General Sales or Services. They also have agreements above 20\% for many level-two and level-three classes.

There is little agreement between Bing Maps and the other data sources. This is mostly due to the small size of the Bing Maps dataset.

\begin{table*}[h]
\setlength{\abovecaptionskip}{0.1cm}
\setlength{\belowcaptionskip}{-0.cm}
  \centering
  \caption{Quantitative comparison with authoritative data}
  \footnotesize
    \begin{tabular}{|c|c|ccc|ccc|ccc|ccc|}
    \hline
    \multicolumn{1}{|c|}{\multirow{2}{1cm}{Class}} & \multirow{2}{1cm}{DataSF} & \multicolumn{3}{c|}{Bing} & \multicolumn{3}{c|}{Google} & \multicolumn{3}{c|}{YP} & \multicolumn{3}{c|}{OSM} \\
\cline{3-14}     &       & Results & Precision & Recall & Results & Precision & Recall & Results & Precision & Recall & Results & Precision & Recall \\
    \hline
    CIE   & 3701  & 38/68 & 0.56  & 0.01  & 606/1413 & 0.43  & 0.16  & 433/978 & 0.44  & 0.12  & 808/1949 & 0.41  & 0.22 \\
    RETAIL/ENT & 4513  & 181/677 & 0.27  & 0.04  & 1295/7274 & 0.17  & 0.28  & 914/4311 & 0.21  & 0.20  & 529/2104 & 0.25  & 0.12 \\
    VISITOR & 508   & 58/167 & 0.35  & 0.11  & 150/586 & 0.26  & 0.30  & 131/397 & 0.33  & 0.26  & 47/53 & 0.89  & 0.09 \\
    MED   & 359   & 1/12  & 0.08  & 0.00  & 151/2055 & 0.07  & 0.42  & 104/1004 & 0.10  & 0.29  & 66/901 & 0.07  & 0.18 \\
    MIPS  & 2138  & 0/3   & 0.00  & 0.00  & 73/339 & 0.22  & 0.03  & 132/499 & 0.26  & 0.06  & 2/7 & 0.29  & 0.00 \\
    RESIDENT & 179028 &       &       &       &       &       &       &       &       &       & 28309/34818 & 0.81  & 0.16 \\
    \hline
    \end{tabular}%
  \label{tab:quan-authoritative}%

\end{table*}%

The agreement between OSM and Google Places or the Yellow Pages is mixed. This agreement can be high for some classes. However, there is a clear mismatch between OSM and the other sources in terms of the class taxonomies. For example, OSM labels a large number of parcels with class 1000 Residence or Accommodation but these are nearly all in subclass 1100 Private Household. In contrast, all of the parcels labeled by Google Places or the Yellow Pages with class 1000 are in subclass 1300 Hotels, Motels, of Other Accommodation Services. This difference reflects the fact that Google Places and the Yellow Pages provide POIs while OSM contains information about residential areas. 

\vspace{-2pt}
\subsection{Comparison with Authoritative Data}

We here compare the data sources with the coarse-grain authoritative data. This requires aligning our LBCS classes with the 10 classes of the authoritative data shown in table \ref{tab:DataSFClasses}. To do this, we assign classes 2100, 5200, and 5300 to RETAIL/ENT; classes 6100 and 6600 to CIE; classes 6200 and 6300 to MIPS; class 1100 to RESIDENT; class 1300 to VISITOR; and class 6500 to MED. Some of the LBCS classes are not assigned due to the mismatch between the two taxonomies. We do not use authoritative classes MIXED and MIXRES due to how broad and ambiguous they are. We also do not use class PDR since our data sources do not cover this class.

Table \ref{tab:quan-authoritative} shows the comparison between each of the data sources and the authoritative data. These results are calculated differently than in table \ref{tab:quan-results} since we treat the authoritative data as the ground truth. The second column shows the counts of parcels labeled with the authoritative data classes. For each data source, we report the number of parcels labeled correctly by that source as well as the precision and recall. For example, 38 of 68 parcels labeled by Bing Maps as CIE are correct according to the authoritative data. This represents a precision of 0.56 and a recall of 0.01.

Google Places and the Yellow Pages are seen to be the best datasets in terms of precision and recall. However, even at this coarse granularity, neither of them achieves precision or recall rates above 0.5 for any class. Bing has very low recall due to its small size. OSM has higher precision than recall and is able to achieve precision above 0.8 for classes VISITOR and RESIDENT. This again emphasizes its difference with the POI data. 


\section{Conclusion and Future Work}

We compared four open-source data sources for fine-grain land-use mapping at the parcel level for San Francisco. We observed limited agreement among the data sources as well as limited accuracy with respect to coarse-grain authoritative data. These results suggest that, at least, the four sources considered are not sufficient for mapping land use over a large geographic region particularly with respect to the proposed fine-grain land-use taxonomy.

This motivates future work on investigating and integrating additional data sources especially ones with dense spatial coverage.


\section{Acknowledgments}
This work was funded in part by a National Science Foundation CAREER grant, \#IIS-1150115.

\bibliographystyle{ACM-Reference-Format}
\bibliography{sample-bibliography} 


\begin{thebibliography}{00}


\ifx \showCODEN    \undefined \def \showCODEN     #1{\unskip}     \fi
\ifx \showDOI      \undefined \def \showDOI       #1{{\tt DOI:}\penalty0{#1}\ }
  \fi
\ifx \showISBNx    \undefined \def \showISBNx     #1{\unskip}     \fi
\ifx \showISBNxiii \undefined \def \showISBNxiii  #1{\unskip}     \fi
\ifx \showISSN     \undefined \def \showISSN      #1{\unskip}     \fi
\ifx \showLCCN     \undefined \def \showLCCN      #1{\unskip}     \fi
\ifx \shownote     \undefined \def \shownote      #1{#1}          \fi
\ifx \showarticletitle \undefined \def \showarticletitle #1{#1}   \fi
\ifx \showURL      \undefined \def \showURL       #1{#1}          \fi
\providecommand\bibfield[2]{#2}
\providecommand\bibinfo[2]{#2}
\providecommand\natexlab[1]{#1}

\bibitem[\protect\citeauthoryear{Adam, Mutanga, Odindi, and Abdel-Rahman}{Adam
  et~al\mbox{.}}{2014}]%
        {Elhadi14}
\bibfield{author}{\bibinfo{person}{Elhadi Adam}, \bibinfo{person}{Onisimo
  Mutanga}, \bibinfo{person}{John Odindi}, {and} \bibinfo{person}{Elfatih~M.
  Abdel-Rahman}.} \bibinfo{year}{2014}\natexlab{}.
\newblock \showarticletitle{Land-use/cover classification in a heterogeneous
  coastal landscape using RapidEye imagery: evaluating the performance of
  random forest and support vector machines classifiers}.
\newblock \bibinfo{journal}{{\em International Journal of Remote Sensing\/}}
  \bibinfo{volume}{{35}, 10} (\bibinfo{year}{2014}),
  \bibinfo{pages}{3440--3458}.
\newblock


\bibitem[\protect\citeauthoryear{Arsanjani, Helbich, Bakillah, Hagenauer, and
  Zipf}{Arsanjani et~al\mbox{.}}{2013}]%
        {Jamal13}
\bibfield{author}{\bibinfo{person}{Jamal~Jokar Arsanjani},
  \bibinfo{person}{Marco Helbich}, \bibinfo{person}{Mohamed Bakillah},
  \bibinfo{person}{Julian Hagenauer}, {and} \bibinfo{person}{Alexander Zipf}.}
  \bibinfo{year}{2013}\natexlab{}.
\newblock \showarticletitle{Toward mapping land-use patterns from volunteered
  geographic information}.
\newblock \bibinfo{journal}{{\em International Journal of Geographical
  Information Science\/}} \bibinfo{volume}{{27}, 12} (\bibinfo{year}{2013}),
  \bibinfo{pages}{2264--2278}.
\newblock


\bibitem[\protect\citeauthoryear{Arsanjani, Mooney, Zipf, and
  Schauss}{Arsanjani et~al\mbox{.}}{2015}]%
        {arsanjani2015quality}
\bibfield{author}{\bibinfo{person}{Jamal~Jokar Arsanjani},
  \bibinfo{person}{Peter Mooney}, \bibinfo{person}{Alexander Zipf}, {and}
  \bibinfo{person}{Anne Schauss}.} \bibinfo{year}{2015}\natexlab{}.
\newblock \showarticletitle{Quality assessment of the contributed land use
  information from OpenStreetMap versus authoritative datasets}.
\newblock In \bibinfo{booktitle}{{\em OpenStreetMap in GIScience}}. Springer,
  \bibinfo{pages}{37--58}.
\newblock


\bibitem[\protect\citeauthoryear{Association}{Association}{2010}]%
        {LBCS}
\bibfield{author}{\bibinfo{person}{American~Planning Association}.}
  \bibinfo{year}{2010}\natexlab{}.
\newblock \bibinfo{title}{Land Based Classification Standards}.
\newblock   (\bibinfo{year}{2010}).
\newblock
\showURL{%
\url{https://www.planning.org/lbcs/}}


\bibitem[\protect\citeauthoryear{Bakillah, Liang, Mobasheri, Arsanjani, and
  Zipf}{Bakillah et~al\mbox{.}}{2014}]%
        {Mohamed14}
\bibfield{author}{\bibinfo{person}{Mohamed Bakillah}, \bibinfo{person}{Steve
  Liang}, \bibinfo{person}{Amin Mobasheri}, \bibinfo{person}{Jamal~Jokar
  Arsanjani}, {and} \bibinfo{person}{Alexander Zipf}.}
  \bibinfo{year}{2014}\natexlab{}.
\newblock \showarticletitle{Fine-resolution population mapping using
  OpenStreetMap points-of-interest}.
\newblock \bibinfo{journal}{{\em International Journal of Geographical
  Information Science\/}} \bibinfo{volume}{{28}, 9} (\bibinfo{year}{2014}),
  \bibinfo{pages}{1940--1963}.
\newblock


\bibitem[\protect\citeauthoryear{Cheng, Han, Guo, Liu, Bu, and Ren}{Cheng
  et~al\mbox{.}}{2015}]%
        {Cheng15}
\bibfield{author}{\bibinfo{person}{G. Cheng}, \bibinfo{person}{J. Han},
  \bibinfo{person}{L. Guo}, \bibinfo{person}{Z. Liu}, \bibinfo{person}{S. Bu},
  {and} \bibinfo{person}{J. Ren}.} \bibinfo{year}{2015}\natexlab{}.
\newblock \showarticletitle{Effective and Efficient Midlevel Visual
  Elements-Oriented Land-Use Classification Using VHR Remote Sensing Images}.
\newblock \bibinfo{journal}{{\em IEEE Transactions on Geoscience and Remote
  Sensing\/}} \bibinfo{volume}{{53}, 8} (\bibinfo{year}{2015}),
  \bibinfo{pages}{4238--4249}.
\newblock


\bibitem[\protect\citeauthoryear{DataSF}{DataSF}{2017}]%
        {SFopendata}
\bibfield{author}{\bibinfo{person}{DataSF}.} \bibinfo{year}{2017}\natexlab{}.
\newblock \bibinfo{title}{Open data: land use}.
\newblock   (\bibinfo{year}{2017}).
\newblock
\showURL{%
\url{https://data.sfgov.org/Housing-and-Buildings/Land-Use/us3s-fp9q/data}}


\bibitem[\protect\citeauthoryear{Goodchild}{Goodchild}{2007}]%
        {goodchild2007citizens}
\bibfield{author}{\bibinfo{person}{Michael~F Goodchild}.}
  \bibinfo{year}{2007}\natexlab{}.
\newblock \showarticletitle{Citizens as sensors: the world of volunteered
  geography}.
\newblock \bibinfo{journal}{{\em GeoJournal\/}} \bibinfo{volume}{{69}, 4}
  (\bibinfo{year}{2007}), \bibinfo{pages}{211--221}.
\newblock


\bibitem[\protect\citeauthoryear{Inc}{Inc}{2017}]%
        {Google}
\bibfield{author}{\bibinfo{person}{Google Inc}.}
  \bibinfo{year}{2017}\natexlab{}.
\newblock \bibinfo{title}{Google Places API}.
\newblock   (\bibinfo{year}{2017}).
\newblock
\showURL{%
\url{https://developers.google.com/places/web-service/search}}


\bibitem[\protect\citeauthoryear{Jiang, Alves, Rodrigues, Ferreira, and
  Pereira}{Jiang et~al\mbox{.}}{2015}]%
        {jiang2015mining}
\bibfield{author}{\bibinfo{person}{Shan Jiang}, \bibinfo{person}{Ana Alves},
  \bibinfo{person}{Filipe Rodrigues}, \bibinfo{person}{Joseph Ferreira}, {and}
  \bibinfo{person}{Francisco~C Pereira}.} \bibinfo{year}{2015}\natexlab{}.
\newblock \showarticletitle{Mining point-of-interest data from social networks
  for urban land use classification and disaggregation}.
\newblock \bibinfo{journal}{{\em Computers, Environment and Urban Systems\/}}
  \bibinfo{volume}{53} (\bibinfo{year}{2015}), \bibinfo{pages}{36--46}.
\newblock


\bibitem[\protect\citeauthoryear{Leung and Newsam}{Leung and Newsam}{2012}]%
        {Leung:2012:EGI:2390790.2390794}
\bibfield{author}{\bibinfo{person}{Daniel Leung} {and} \bibinfo{person}{Shawn
  Newsam}.} \bibinfo{year}{2012}\natexlab{}.
\newblock \showarticletitle{Exploring Geotagged Images for Land-use
  Classification}. In \bibinfo{booktitle}{{\em Proceedings of the ACM
  Multimedia 2012 Workshop on Geotagging and Its Applications in Multimedia}}
  \bibinfo{series}{{\em (GeoMM '12)}}. \bibinfo{pages}{3--8}.
\newblock
\showISBNx{978-1-4503-1590-6}


\bibitem[\protect\citeauthoryear{Liu, He, Yao, Zhang, Liang, Wang, and
  Hong}{Liu et~al\mbox{.}}{2017}]%
        {Liu17}
\bibfield{author}{\bibinfo{person}{Xiaoping Liu}, \bibinfo{person}{Jialv He},
  \bibinfo{person}{Yao Yao}, \bibinfo{person}{Jinbao Zhang},
  \bibinfo{person}{Haolin Liang}, \bibinfo{person}{Huan Wang}, {and}
  \bibinfo{person}{Ye Hong}.} \bibinfo{year}{2017}\natexlab{}.
\newblock \showarticletitle{Classifying urban land use by integrating remote
  sensing and social media data}.
\newblock \bibinfo{journal}{{\em International Journal of Geographical
  Information Science\/}} \bibinfo{volume}{{31}, 8} (\bibinfo{year}{2017}),
  \bibinfo{pages}{1675--1696}.
\newblock


\bibitem[\protect\citeauthoryear{Manandhar, Odeh, and Ancev}{Manandhar
  et~al\mbox{.}}{2009}]%
        {rs1030330}
\bibfield{author}{\bibinfo{person}{Ramita Manandhar}, \bibinfo{person}{Inakwu
  O.~A. Odeh}, {and} \bibinfo{person}{Tiho Ancev}.}
  \bibinfo{year}{2009}\natexlab{}.
\newblock \showarticletitle{Improving the Accuracy of Land Use and Land Cover
  Classification of Landsat Data Using Post-Classification Enhancement}.
\newblock \bibinfo{journal}{{\em Remote Sensing\/}} \bibinfo{volume}{{1}, 3}
  (\bibinfo{year}{2009}), \bibinfo{pages}{330--344}.
\newblock


\bibitem[\protect\citeauthoryear{Mao, Thakur, and Bhaduri}{Mao
  et~al\mbox{.}}{2016}]%
        {Mao:2016:EMP:3007540.3007549}
\bibfield{author}{\bibinfo{person}{Huina Mao}, \bibinfo{person}{Gautam Thakur},
  {and} \bibinfo{person}{Budhendra Bhaduri}.} \bibinfo{year}{2016}\natexlab{}.
\newblock \showarticletitle{Exploiting Mobile Phone Data for Multi-category
  Land Use Classification in Africa}. In \bibinfo{booktitle}{{\em Proceedings
  of the 2Nd ACM SIGSPATIAL Workshop on Smart Cities and Urban Analytics}}
  \bibinfo{series}{{\em (UrbanGIS '16)}}. Article \bibinfo{articleno}{9}, 6
  pages.
\newblock
\showISBNx{978-1-4503-4583-5}


\bibitem[\protect\citeauthoryear{Microsoft}{Microsoft}{2017}]%
        {Bing}
\bibfield{author}{\bibinfo{person}{Microsoft}.}
  \bibinfo{year}{2017}\natexlab{}.
\newblock \bibinfo{title}{Bing Maps API}.
\newblock   (\bibinfo{year}{2017}).
\newblock
\showURL{%
\url{https://msdn.microsoft.com/en-us/library/gg585126.aspx}}


\bibitem[\protect\citeauthoryear{Saadat, Adamowski, Bonnell, Sharifi, Namdar,
  and Ale-Ebrahim"}{Saadat et~al\mbox{.}}{2011}]%
        {SAADAT2011608}
\bibfield{author}{\bibinfo{person}{"Hossein Saadat}, \bibinfo{person}{Jan
  Adamowski}, \bibinfo{person}{Robert Bonnell}, \bibinfo{person}{Forood
  Sharifi}, \bibinfo{person}{Mohammad Namdar}, {and} \bibinfo{person}{Sasan
  Ale-Ebrahim"}.} \bibinfo{year}{2011}\natexlab{}.
\newblock \showarticletitle{Land use and land cover classification over a large
  area in Iran based on single date analysis of satellite imagery}.
\newblock \bibinfo{journal}{{\em ISPRS Journal of Photogrammetry and Remote
  Sensing\/}} \bibinfo{volume}{{66}, 5} (\bibinfo{year}{2011}),
  \bibinfo{pages}{608 -- 619}.
\newblock
\showISSN{0924-2716}


\bibitem[\protect\citeauthoryear{Toole, Ulm, Gonz\'{a}lez, and Bauer}{Toole
  et~al\mbox{.}}{2012}]%
        {Toole12}
\bibfield{author}{\bibinfo{person}{Jameson~L. Toole}, \bibinfo{person}{Michael
  Ulm}, \bibinfo{person}{Marta~C. Gonz\'{a}lez}, {and} \bibinfo{person}{Dietmar
  Bauer}.} \bibinfo{year}{2012}\natexlab{}.
\newblock \showarticletitle{Inferring Land Use from Mobile Phone Activity}. In
  \bibinfo{booktitle}{{\em Proceedings of the ACM SIGKDD International Workshop
  on Urban Computing}} \bibinfo{series}{{\em (UrbComp '12)}}.
  \bibinfo{pages}{1--8}.
\newblock
\showISBNx{978-1-4503-1542-5}


\bibitem[\protect\citeauthoryear{Yao, Li, Liu, Liu, Liang, Zhang, and Mai}{Yao
  et~al\mbox{.}}{2017}]%
        {Yao17}
\bibfield{author}{\bibinfo{person}{Yao Yao}, \bibinfo{person}{Xia Li},
  \bibinfo{person}{Xiaoping Liu}, \bibinfo{person}{Penghua Liu},
  \bibinfo{person}{Zhaotang Liang}, \bibinfo{person}{Jinbao Zhang}, {and}
  \bibinfo{person}{Ke Mai}.} \bibinfo{year}{2017}\natexlab{}.
\newblock \showarticletitle{Sensing spatial distribution of urban land use by
  integrating points-of-interest and Google Word2Vec model}.
\newblock \bibinfo{journal}{{\em International Journal of Geographical
  Information Science\/}} \bibinfo{volume}{{31}, 4} (\bibinfo{year}{2017}),
  \bibinfo{pages}{825--848}.
\newblock


\bibitem[\protect\citeauthoryear{Zheng and Zheng}{Zheng and Zheng}{2014}]%
        {Zheng2014}
\bibfield{author}{\bibinfo{person}{Shudan Zheng} {and}
  \bibinfo{person}{Jianghua Zheng}.} \bibinfo{year}{2014}\natexlab{}.
\newblock \showarticletitle{Assessing the Completeness and Positional Accuracy
  of OpenStreetMap in China}.
\newblock In \bibinfo{booktitle}{{\em Thematic Cartography for the Society}}.
  Springer International Publishing, Cham, \bibinfo{pages}{171--189}.
\newblock


\bibitem[\protect\citeauthoryear{Zhu and Newsam}{Zhu and Newsam}{2015}]%
        {Zhu:2015:LUC:2820783.2820851}
\bibfield{author}{\bibinfo{person}{Yi Zhu} {and} \bibinfo{person}{Shawn
  Newsam}.} \bibinfo{year}{2015}\natexlab{}.
\newblock \showarticletitle{Land Use Classification Using Convolutional Neural
  Networks Applied to Ground-level Images}. In \bibinfo{booktitle}{{\em
  Proceedings of the 23rd SIGSPATIAL International Conference on Advances in
  Geographic Information Systems}} \bibinfo{series}{{\em (SIGSPATIAL '15)}}.
  Article \bibinfo{articleno}{61}, 4 pages.
\newblock
\showISBNx{978-1-4503-3967-4}


\bibitem[\protect\citeauthoryear{Zielstra, Hochmair, and Neis}{Zielstra
  et~al\mbox{.}}{2013}]%
        {TGIS:TGIS12037}
\bibfield{author}{\bibinfo{person}{Dennis Zielstra},
  \bibinfo{person}{Hartwig~H. Hochmair}, {and} \bibinfo{person}{Pascal Neis}.}
  \bibinfo{year}{2013}\natexlab{}.
\newblock \showarticletitle{Assessing the Effect of Data Imports on the
  Completeness of OpenStreetMap - A United States Case Study}.
\newblock \bibinfo{journal}{{\em Transactions in GIS\/}} \bibinfo{volume}{{17},
  3} (\bibinfo{year}{2013}), \bibinfo{pages}{315--334}.
\newblock


\end{thebibliography}

\begin{landscape}

\begin{table}[htbp]
\setlength{\abovecaptionskip}{0.3cm}
\setlength{\belowcaptionskip}{-0.cm}
  \centering
  \caption{Quantitative inter-dataset comparisons}
    \begin{tabular}{|p{0.8cm}|p{0.8cm}p{0.8cm}p{0.8cm}p{0.8cm}|p{1.8cm}p{2cm}p{1.8cm}p{1.8cm}p{1.8cm}p{1.8cm}p{1.5cm}|}
    \hline
          & \multicolumn{4}{c|}{Mapping results} & \multicolumn{6}{c}{Comparison results} &  \\
\cline{2-12}    Class &Google & Bing & YP & OSM & Google\&Bing & Google\&YP & Bing\&YP & OSM\&YP &Google\&OSM & Bing\&OSM & Google \& Bing \& OSM \&YP  \\
    \hline
    1000  & 586   & 167   & 397   & 34871 & {113 (17.66\%)} & 267 (37.29\%) & {96 (20.51\%)} & {16 (0.05\%)} & 28(0.08\%)    & 7 (0.02\%)     & 2 \\
    1100  &       &       &       & 34818 &       &  &       &       &       &       &  \\
    1300  & 586   & 167   & 397   & 53    & {113 (17.66\%)} & 267 (37.29\%) &{96 (20.51\%)} & 7 (1.58\%)      & 6 (0.95\%)  & 1 (0.46\%)     &0  \\
    2000  & 10481 & 1617  & 8661  & 10059 & {1375 (12.82\%)} & 6323 (49.33\%) & {1317 (14.70)} & {3941 (26.67\%)} & {4732 (29.93\%)} & {1067 (10.06\%)} & 871 \\
    2100  & 5996  & 627   & 3256  & 1898  &{474 (7.71\%)} & 2271 (32.53\%) &{375 (10.69\%)} &{829 (19.17\%)} & {1254 (18.89\%)} & {202 (8.70\%)} & 143 \\
    2110  & 799   & 10    & 586   &       &{5 (0.62\%)} & 341 (32.66\%) &{5 (0.85\%)} &       &       &       &  \\
    2120  & 989   & 93    & 362   &       & {34 (3.24\%)} & 154 (12.87\%) & {21 (4.84\%)} &       &       &       &  \\
    2130  & 1640  & 259   & 1095  &       &{153 (8.76\%)} & 587 (27.33\%) &{129 (10.53\%)} &       &       &       &  \\
    2140  & 156   &       & 151   &       &       & 61 (24.80\%) &       &       &       &       &  \\
    2150  & 814   & 254   & 1252  &       & {50 (4.91\%)} & 452 (28.00\%) &{138 (10.09\%)} &       &       &       &  \\
    2160  & 213   & 40    & 120   &       & {24 (10.48\%)} & 83 (33.20\%) &{22 (15.94\%)} &       &       &       &  \\
    2200  & 808   & 135   & 1426  & 352   &{73 (8.39\%)} & 494 (28.39\%) & {97 (6.63\%)} &{182 (11.40\%)} & {189 (19.46\%)} & {56 (12.99\%)} & 40 \\
    2300  & 1124  & 9     & 1101  & 32    & {5 (0.44\%)} & 471 (26.85\%) & {5 (0.45\%)} & {11 (0.98\%)} & {12 (1.05\%)} & {3 (7.89\%)} & 3 \\
    2400  & 1676  & 17    & 1948  & 132   &{13 (0.77\%)} & 727 (25.09\%) & {2 (0.10\%)} & {34 (1.66\%)} &{67 (3.85\%)} & {9 (6.43\%)} & 0 \\
    2500  & 3590  & 1004  & 3045  & 2220  & {781 (20.48\%)} & 2273(52.11\%) &{741 (22.40\%)} & {1400 (36.22\%)} & {1718 (41.98\%)} & {529 (19.63\%)} & 393 \\
    2600  & 1858  &       & 1630  &       &       & 970 (38.52\%) &       &       &       &       &  \\
    2700  & 52    &       & 102   &       &       & 25 (19.38\%) &       &       &       &       &  \\
    \hline
    4000  & 1898  & 102   & 335   & 356   & {44 (2.25\%)} & 148 (7.10\%) & {42 (10.63\%)} & {48 (7.47\%)} &{99 (4.59\%)} & {17 (3.85\%)} & 11 \\
    4100  & 1808  & 86    & 271   & 294   & {37 (1.99\%)} & 112 (5.69\%) & {33 (10.19\%)} & {22 (4.05\%)} & {62 (3.04\%} & {12 (3.26\%)} & 6 \\
    4200  & 90    & 16    & 68    & 55    &{6 (6.00\%)} & 34 (27.42\%) & {8 (10.53\%)} &{18 (17.14\%)} &{27 (22.88\%)} & {5 (7.58\%)} & 5 \\
    \hline
    5000  & 1167  & 50    & 964   & 206   &{15 (1.25\%)} & 369 (20.94\%) & {17 (1.71\%)} & {55 (4.93\%)} & {88 (6.85\%)} & {4 (1.59\%)} & 0 \\
    5100  & 540   & 13    & 455   & 106   &{2 (0.36\%)} & 181 (22.24\%) &{4 (0.86\%)} &{17 (3.13\%)} & {30 (4.87\%)} & {4 (3.48\%)} & 0 \\
    5200  & 131   & 24    & 91    & 8     & {6 (4.03\%)} & {25 (12.69\%)} &{10 (9.52\%)} & {0 (0)} & {0 (0)}  &{0 (0)} & 0 \\
    5300  & 569   & 14    & 505   & 91    & {1 (0.17\%)} & 151 (16.36\%) & {0 (0)} & {30 (5.30\%)} & {47 (7.67\%)} & {0 (0)} & 0 \\
    5400  & 2     & 0     & 5     &       &{0 (0)} & 0 (0) &{0 (0)} &       &       &       &  \\
    \hline
    6000  & 3536  & 86    & 2218  & 2910  & {58 (1.63\%)} & 1303 (29.27\%) &{44 (1.95\%)} &{447 (9.55\%)} & {766 (13.49\%)} & {42 (1.42\%)} & 22 \\
    6100  & 862   & 68    & 501   & 1450  & {41 (4.61\%)} & 228 (20.09\%) & {36 (6.75\%)} & {105 (5.69\%)} & {246 (11.91\%)} & {25 (1.67\%)} & 13 \\
    6200  & 282   & 3     & 449   & 7     & {0 (0)} & 106 (19.69\%) & {0 (0)} & {2 (0.44\%)} & {3 (1.05\%)} & {0 (0)} & 0 \\
    6300  & 57    &       & 50    &       &       & 29 (37.18\%) &       &       &       &       &  \\
    6400  & 79    & 3     & 53    & 96    & {2 (2.50\%)} & 18 (15.79\%) & {2 (3.70\%)} &{16 (12.03\%)} & {42 (31.58\%)} & {2 (2.06\%)} & 2 \\
    6500  & 2055  & 12    & 1004  & 901   &{9 (0.44\%)} & 631 (25.99\%) &{3 (0.30\%)} & {138 (7.81\%)} & {215 (7.84\%)} & {8 (0.88\%)} & 3 \\
    6600  & 551   &       & 477   & 499   &       & 273 (36.16\%) &       & {137 (16.33\%)} & {192 (22.38\%)} &       &  \\
    6700  & 26    &       & 30    &       &       & 14 (33.33\%) &       &       &       &       &  \\
    \hline
    \end{tabular}%
  \label{tab:quan-results}%
\end{table}%
\end{landscape}

\begin{table*}
\setlength{\abovecaptionskip}{0.2cm}
\setlength{\belowcaptionskip}{-0.2cm}
\caption{Selected LBCS classes and descriptions, along with POI taxonomic assignments (part 1 of 2)}
\begin{tabular}{|p{0.1\columnwidth}|p{2cm}|p{1cm}|p{3cm}|p{0.3\columnwidth}|p{2.5cm}|p{2.5cm}|} 
\hline
Level-1 class & Functions & Level-2 class & Functions&Google places type& Bing entity type& YellowPage keyword \\ 
\hline 
\multirow{2}{1cm}{1000} & Residence or accommodation functions  &1100 & Private household & & & \\
\cline{3-7}
 & & 1300 &  Hotels, motels, or other accommodation services&lodging&Hotel &lodging\\
\hline
\multirow{7}{1cm}{2000} & \multirow{7}{2cm}{General sales or services} & 2100 & Retail sales or service & \multicolumn{3}{c|}{more details in table \ref{tab:LBCS2}}\\
\cline{3-7}
&& 2200 & Finance and Insurance &bank, insurance_agency, atm  & ATM, Bank & bank, insurance agency, ATM\\
\cline{3-7}
& & 2300  & Real estate, and rental and leasing& car_rental, movie_rental, real_estate_agency& Rental Car Agency&  car rental, movie rental, real estate agency\\
\cline{3-7}
&& 2400 & Business, professional, scientific, and technical services&lawyer, post_office, travel_agency, veterinary_care, accounting& Tourist Information, Post Office&lawyer, post office, travel agency, veterinary care, accounting\\
\cline{3-7}
&& 2500 & Food services&restaurant, cafe, night_club, bar& Restaurant & restaurant, caf\'e, night club, bar\\
\cline{3-7}
&& 2600 & Personal services&laundry, spa, hair_care, beauty_salon&&laundry, spa, hair care, beauty salon\\
\cline{3-7}
&& 2700 & Pet and animal sales or service (except veterinary)&pet_store& &pet store\\
\hline
\multirow{4}{1cm}{4000}&\multirow{4}{2cm}{Transportation, communication, information, and utilities} & 4100 & Transportation services &bus_station, subway_station, taxi_stand, transit_station, parking &Bus Station, Commuter Rail Station, Parking Garage or House, Transportation Service&bus station, subway station, taxi stand,transit station, parking\\
\cline{3-7}
&&  4200 & Communications and information&library& Library&library\\
\hline
\multirow{4}{1cm}{5000}&\multirow{4}{2cm}{Arts, entertainment, and recreation} &5100 & Performing arts or supporting  establishment&art_gallery, movie_theater, stadium & Cinema, Performing Arts&art gallery, movie theater, stadium \\
\cline{3-7}
&& 5200&Museums and other special purpose recreational institutions&aquarium, zoo, museum & Animal Park, Historical Monument&aquarium, zoo, museum \\
\cline{3-7}
&&5300& Amusement, sports, or recreation establishment&park, amusement_park, casino, gym, bowling_alley &Park or Recreation Area& park, amusement park, casino, gym, bowling alley \\
\cline{3-7}
&&5400 & Camps, camping, and related establishments&campground&Campground&campground\\
\hline
\multirow{7}{1cm}{6000}&\multirow{7}{2cm}{Education, public admin., health care, and other inst.} & 6100 & Educational services&school, university & Higher Education, School &school, university\\
\cline{3-7}
&&6200 & Public administration&city_hall, courthouse, local_government_office& Civic or Community Centre, Convention or Exhibition Centre, City Hall, Court House &city hall, courthouse, local government office\\
\cline{3-7}
&&6300 & Other government functions&embassy& & embassy\\
\cline{3-7}
&&6400 & Public Safety&fire_station, police& Police Station &fire station, police\\
\cline{3-7}
&&6500 & Health and human services&dentist, hospital, doctor&Hospital&dentist, hospital, doctor\\
\cline{3-7}
&&6600 & Religious institutions&church, hindu_temple, mosque, synagogue&&church, hindu temple, mosque, synagogue\\
\cline{3-7}
&&6700 & Death care services6700&funeral_home&&funeral home\\
\hline
\end{tabular} 
\label{tab:LBCS1}
\end{table*}

\begin{table*}
\setlength{\abovecaptionskip}{0.2cm}
\setlength{\belowcaptionskip}{-0.2cm}
  \caption{Selected LBCS classes and descriptions, along with POI taxonomic assignments (part 2 of 2)}
    \begin{tabular}{|c|p{3cm}|p{3cm}|p{3cm}|p{3cm}|}
    \hline
    Level-3 class & Description & Google place type & Bing entity type & YellowPages keywords \\
    \hline
    2110  & Automobile sales or service establishment & car\_dealer, car\_repair, car\_wash, bicycle\_store,  gas\_station & Auto Dealerships, Petrol or Gasoline Station, Motorcycle Dealership & car dealer, car repair, car wash, bicycle store,  gas station \\
    \hline
    2120  & Heavy consumer goods sales or service & department\_store, furniture\_store, hardware\_store, home\_goods\_store & Department Store, Home Specialty Store, Home Improvement \& Hardware Store & department store, furniture store, hardware store, home goods store \\
    \hline
    2130  & Durable consumer goods sales and service & book\_store, clothing\_store, electronics\_store, jewelry\_store, shoe\_store & Book Store, Consumer Electronics Store, Clothing Store, Sporting Goods Store & book store, clothing store, electronics store, jewelry store, shoe store \\
    \hline
    2140  & Consumer goods, other & florist & & florist \\
    \hline
    2150  & Grocery, food, beverage, dairy, etc. & bakery, convenience\_store, liquor\_store & Grocery Store, Convenience Store, Coffee Shop & bakery, convenience store, liquor store \\
    \hline
    2160  & Health and personal care & pharmacy & Pharmacy & pharmacy \\
    \hline
    \end{tabular}%
  \label{tab:LBCS2}%
\end{table*}

\end{document}